\documentclass[aps,amsmath,amssymb,prb]{revtex4}
\usepackage{graphicx}
\usepackage{latexsym}
\usepackage{amssymb}

\newcommand{\BE}{\begin{equation}}
\newcommand{\EE}{\end{equation}}
\newcommand{\BA}{\begin{eqnarray}}
\newcommand{\EA}{\end{eqnarray}}
\usepackage{graphicx}
\usepackage{dcolumn}
\usepackage{bm}
\input epsf.tex

\def\mn{{\medskip\par}}

\def\sn{{\smallskip\par\noindent}}

\begin{document}





\begin{center}
\begin{Large}
{\bf\textsf{Light deflection by photonic crystals}}
\end{Large}

\mn

{\bf\textsf{Chao-Hsien Kuo and Zhen Ye}}

\sn {\scriptsize\sl Wave Phenomena Laboratory, Department of
Physics, National Central University, Chungli, Taiwan 320,
Republic of China.}
\end{center}


\begin{quotation}
\noindent {\bf When propagating through periodically structured
media, i.~e. photonic crystals, optical waves will be modulated
with the periodicity. As a result, the dispersion of waves will no
longer behave as in a free space, and so called frequency band
structures appear. Under certain conditions, waves may be
prohibited from propagation in certain or all directions,
corresponding to partial and complete bandgaps respectively. Here
we report a new fascinating phenomenon associated with the partial
gaps, that is, deflection of optical waves. This phenomenon will
render novel applications in manipulating light flows.}
\end{quotation}

\mn The band phenomenon was first investigated for electrons in
solids. The well-known Bloch theorem has been proposed and led to
successful explanation of some important properties of solids such
as conducting, semi-conducting, and insulating states. Applying
these concepts to electromagnetic waves started and flourished
with the seminal papers published in 1987 \cite{Yab}. Not only all
the phenomena previously observed or discussed for electronic
systems are successfully transplanted to classical systems, but
many more significant and novel ideas, and applications have well
gone beyond expectation, and are so far reaching that a fruitful
new field called photonic crystals has come into existence,
signified by the establishment of a comprehensive webpage
\cite{web}.

\mn Photonic crystals (PCs) provide a new possibility in designing
optoelectronic devices. Up to date, most applications have mainly
relied on the existence of complete bandgaps. This includes, for
example, cavity laser, optical fibers, and optical waveguides
\cite{Opto}. Here we demonstrate a new property of PCs, which is
associated with partial bandgaps, referring to the situation in
which waves are prohibited from propagation along certain
directions.

\mn We show that partial bandgaps can give rise to a fascinating
phenomenon: deflection of optical waves. This phenomenon would
allow for better optical steering, bending and beam collimating.
It may also help explain some recent mysteries about the negative
refraction or left-handed-material (LHM) behavior and the
amphoteric refraction revealed by PCs \cite{PRB,exp,Chuang,Zhang}.

\mn The systems considered here are two dimensional photonic
crystals made of arrays of parallel dielectric cylinders placed in
a uniform medium, which we assume to be air. Such systems are
common in both theoretical simulations or experimental
measurements of two dimensional PCs\cite{PRB,McCall}. For brevity,
we only consider the E-polarized waves, that is, the electric
field is kept parallel to the cylinders. The following parameters
are used in the simulation. (1) The dielectric constant of the
cylinders is 14, and the cylinders are arranged to form a square
lattice. (2) The lattice constant is $a$ and the radius of the
cylinders is 0.3$a$; in the computation, all lengths are scaled by
the lattice constant. (3) The unit for frequency is $2\pi c/a$.

\mn The band structure is plotted in Fig.~1, and the qualitative
features are similar to that obtained for a square array of
alumina rods in air. A complete band gap is shown between
frequencies of 0.22 and 0.28. Just below the complete gap, there
is a regime, sandwiched by two horizontal lines, of partial band
gap in which waves are not allowed to travel along the $\Gamma X$
or [10] direction.

\mn We wish to examine the properties of the energy flow of the
eigenmodes which correspond to the frequency bands shown in
Fig.~1. Differing from the common approaches which mainly rely on
the curvatures of frequency bands to infer the energy flow, we
consider the genuine definition of the energy flow. By Bloch's
theorem, the eigenmodes corresponding to the frequency bands of
PCs can be expressed as $E_{\vec{K}}(\vec{r}) =
e^{i\vec{K}\cdot\vec{r}}u_{\vec{K}}(\vec{r})$, where $\vec{K}$ is
the Bloch vector, as the wave vector, and $u_{\vec{K}}(\vec{r})$
is a periodic function with the periodicity of the lattice. When
expressing $E_{\vec{K}}(\vec{r})$ as
$|E_{\vec{K}}(\vec{r})|e^{i\theta_{\vec{K}}(\vec{r})}$, the
corresponding energy flow is derived as
$\vec{J}_{\vec{K}}(\vec{r}) \propto
|E_{\vec{K}}(\vec{r})|^2\nabla\theta_{\vec{K}}(\vec{r})$; clearly
$\theta_{\vec{K}}$ combines the phase from the term
$e^{i\vec{K}\cdot\vec{r}}$ and the phase from the function
$u_{\vec{K}}(\vec{r})$, making the determination of the phase and
group velocities in PCs less obvious, or even perhaps not so
useful. To explore the characteristics of the partial bandgap, we
have computed the eigen-field $E_{\vec{K}}(\vec{r})$ and also the
energy flow of the eigenmodes. The results are shown in Fig.~2.

\mn A significant discovery from Fig.~2 is that the energy flow of
eigenmodes does not always follow the direction of the Bloch
vectors. As evidenced by Fig.~2 (b2), the energy tends to flow
into the direction of $\Gamma M$, i.~e the [11] direction, while
the Bloch vector points to an angle of 22.5$^o$ that lies exactly
between $\Gamma X$ and $\Gamma M$. This feature, however, cannot
be inferred from the field pattern in (b1), i.~e. the main wave
front is more or less perpendicular to the direction of the Bloch
vector rather than that of the energy flow, implying that it is
not appropriate to determine the energy flow purely from the field
pattern. For the other two Bloch vectors along $\Gamma X$ and
$\Gamma M$ respectively, the net energy flows and the normals of
the wave fronts follow the directions of the Bloch vectors, as
shown by Figs.~2(a) and (c). From (a1), (b1) and (c1), we see that
the eigenmodes have also periodic structures perpendicular to the
direction of the propagation. The feature in (b2) can be explained
in the context of partial bandgaps. In Fig.~2 (b), the frequency
0.185 lies within the partial bandgap in the $\Gamma X$ direction.
Since waves are not allowed to propagate along $\Gamma X$ and the
directions perpendicular to $\Gamma X$, we may expect from the
symmetry consideration that the energy flow will be tilted towards
$\Gamma M$ for waves whose Bloch vectors make an angle to $\Gamma
X$. Note here that in Fig.~2, the frequency corresponding to the
Bloch vector along $\Gamma X$ is outside the partial gap.

\mn The deviation of the energy flow from the direction of the
Bloch vector, serving as the wave vector, suggests an intriguing
phenomenon, that is, the light deflection. Consider a plane wave
propagation in a uniform medium. When it encounters an
perpendicular interface of PCs with the normal of the interface
being along the direction exactly between $\Gamma X$ and $\Gamma
M$, what can be expected to happen? In the conventional thought,
the wave would keep going straight, and the wave vectors on both
sides of the interface will point to the same direction, as there
are no components in the tangent direction. However, according to
Fig.~2, the flow of the energy will be deflected into the
direction of $\Gamma M$. This is a surprising finding. To confirm
this, we have carried out further simulations. For example, first,
when we chose frequencies that lie outside the partial gap region,
the deflection phenomenon disappears.

\mn We have also studied the transmission properties. In Fig.~3,
we plot the transmission of EM waves across two slabs of photonic
crystals with two different lattice orientations: one is along the
diagonal direction $\Gamma M$, and the other is along the
direction making an equal angle to the $\Gamma X$ and $\Gamma M$
directions. The incidence is normal to the slab interface. In the
simulation, we have employed the standard multiple scattering
method \cite{MST} to compute the transmitted intensity, which is
usually the quantity to be measured in experiments. The
frequencies chosen correspond to those in Fig.~2. We have used
three types of sources in the simulations: (1) plane waves; (2)
collimated waves by guiding the wave propagation through a window
before incidence on the slabs; (3) a line source. The results from
these three scenarios are similar. Here, we see that when the
incidence is along the $\Gamma M$ direction, the transmission
follows a straight path inside the slab. For the second case in
Fig.~3, the transmission direction inside the slab is deflected
towards the $\Gamma M$ direction. Fig.~3 clearly indicates that
there are some favorable directions for waves to travel, fully in
agreement with the above light deflection picture and in
consistency with the results in Fig.~2. We also note that there is
relatively small transmission along the [1-1] direction; by
symmetry, this direction is also a probable passing path for the
waves.

\mn The light deflection in the presence of partial bandgaps can
be important in a number of occasions. First, it may help
designing novel optoelectronic devices in controlling optical
flows. Compared to applications based on complete bandgaps, there
is no reflection loss because the light deflection occurs in the
passing bands, allowing for efficient wave transmission. Second,
the mechanism of the light deflection discussed here may help
understand the recent puzzles about the negative refraction
\cite{PRB} or LHM behavior\cite{Chuang} and the amphoteric
refraction revealed by PCs \cite{Zhang}. To elaborate, we consider
the slab shown in Fig.~3 (b). Assume that a plane wave is incident
on the slab with a tilt angle. We can expect from the above
results that as long as the tilt angle is not too large, the
energy flow will be deflected towards the direction of $\Gamma M$,
no matter the incidence is positive or negative. Consequently, an
amphoteric refraction may prevail and is conceptually illustrated
by Fig.~4.


\sn

\vskip 12pt
\renewcommand{\refname}{}
\makeatletter
\renewcommand{\@biblabel}[1]{\hfill#1.}
\makeatother \vspace{-1.2cm}

\begin{scriptsize}

{Correspondence and requests for materials should be addressed to
Z.Y. (e-mail: zhen@shaw.ca).}

\end{scriptsize}


\newpage

\begin{center}
{\bf \sf Figure Captions}
\end{center}

\begin{description}

\item[Figure 1] {\sf The band structure of a square lattice of
dielectric cylinders. The lattice constant is $a$ and the radius
of the cylinders is $0.3a$. $\Gamma M$ and $\Gamma X$ denote the
[11] and [10] directions respectively. A partial gap is between
the two horizontal lines.}

\item[Figure 2] {\sf Left panel: the field pattern of eigenmodes.
Right panel: the energy flow of the eigenmodes. The eigenmodes
along three directions are considered: (a) $\vec{K} = (0.9\pi/a,
0)$, i.~e. along $\Gamma X$; the corresponding frequency is 0.175;
(b) $\vec{K} = (0.9\pi/a, 0.37\pi/a)$, i.~e. along an angle of
22.5$^o$ exactly between $\Gamma X$ and $\Gamma M$ directions; the
corresponding frequency is 0.185; (c) $\vec{K} = (0.7\pi/a,
0.7\pi/a)$, i.~e. along $\Gamma M$; the corresponding frequency is
0.192. The direction of the Bloch vectors are denoted by the blue
arrows, while the red arrows denote the local energy flow
including the direction and the magnitude. The circles refer to
the cylinders. Both frequencies in (b) and (c) lie within the
partial gap. Due to the periodicity, we only plot the energy flow
within one unit cell. Note that although the features shown by (b)
also hold for other Bloch vectors for which the corresponding
frequencies lie within the partial gap regime, we only plot here
for the case of 22.5$^o$}

\item[Figure 3] {\sf The imaging fields for slabs of photonic
crystal structure. Two lattice orientations are considered: (a)
The slab measures as about 56$\times 10$ and the incidence is
along the [11] direction; (2) the slab measures as 50$\times 13$
and the incidence is along the direction that makes an equal angle
to the [11] and [10] directions. The main lobes in the transmitted
intensities are shown. There is a gap along the [10] direction.}

\item[Figure4] {\sf Conceptual illustration of amphoteric
refraction due to partial bandgaps, or perhaps better termed as
amphoteric deflection. Depending on the incident angle, the
deflection may not exactly follow the direction of $[11]$.}

\end{description}

\end{document}